\documentclass[aps,amsmath,amssymb,twocolumn]{revtex4-1}

\usepackage{graphicx}
\usepackage{dcolumn}
\usepackage{bm}
\usepackage{color}

\begin{document}

\title{Enhancing Near-Field Radiative Heat Transfer with Si-based Metasurfaces}

\author{V. Fern\'andez-Hurtado$^{1,2,3}$}
\author{F. J. Garc\'{\i}a-Vidal$^{1,4}$}
\author{Shanhui Fan$^{2}$}
\author{J. C. Cuevas$^{1,3}$}

\affiliation{$^1$Departamento de F\'{\i}sica Te\'orica de la Materia Condensada and Condensed Matter 
Physics Center (IFIMAC), Universidad Aut\'onoma de Madrid, E-28049 Madrid, Spain}
\affiliation{$^{2}$Department of Electrical Engineering, and Ginzton Laboratory, Stanford University, 
Stanford, California 94305, USA}
\affiliation{$^{3}$Department of Physics, University of Konstanz, D-78457 Konstanz, Germany}
\affiliation{$^{4}$Donostia International Physics Center (DIPC), Donostia/San Sebasti\'an 20018, Spain}

\date{\today}

\begin{abstract}
We demonstrate in this work that the use of metasurfaces provides a viable strategy to largely 
tune and enhance near-field radiative heat transfer between extended structures. In particular, using 
a rigorous coupled wave analysis, we predict that Si-based metasurfaces featuring two-dimensional periodic
arrays of holes can exhibit a room-temperature near-field radiative heat conductance much larger than any 
unstructured material to date. We show that this enhancement, which takes place in a broad range of 
separations, relies on the possibility to largely tune the properties of the surface plasmon polaritons 
that dominate the radiative heat transfer in the near-field regime.
\end{abstract}

\maketitle

Thermal radiation is one of the most ubiquitous physical phenomena. In recent years, there has been a 
renewed interest in this topic due to the confirmation of the long-standing prediction that radiative 
heat transfer can be drastically enhanced for bodies separated by small gaps \cite{Polder1971,Rytov1953}.
This enhancement, which occurs when the gaps are smaller than the thermal wavelength (9.6 $\mu$m at room 
temperature), is due to the contribution of evanescent waves that dominate the near-field regime. The fact that 
this near-field radiative heat transfer (NFRHT) between closely spaced bodies can overcome the far-field 
limit set by the Stefan-Boltzmann law for black bodies has now been verified in a variety of experiments 
exploring different materials, geometrical shapes, and gaps ranging from micrometers to a few nanometers \cite{Kittel2005,Rousseau2009,Shen2009,Ottens2011,Kralik2012,Zwol2012a,Zwol2012b,Guha2012,Worbes2013,
St-Gelais2014,Song2015a,Kim2015,St-Gelais2016,Song2016,Bernardi2016}. These experiments have also triggered 
off the hope that NFRHT could have an impact in different thermal technologies \cite{Song2015b} such 
as thermophotovoltaics \cite{Lenert2014}, heat-assisted magnetic recording \cite{Challener2009,Stipe2010}, 
scanning thermal microscopy \cite{Wilde2006,Kittel2008,Jones2013}, nanolithography \cite{Pendry1999}, 
thermal management \cite{Otey2010,Ben-Abdallah2014} or coherent thermal sources \cite{Carminati1999,Greffet2002}.

In this context, the question on the fundamental limits of NFRHT is attracting a lot of attention
\cite{Miller2015}. So far, the largest NFRHT enhancements in extended structures have been reported
(both theoretically and experimentally) for polar dielectrics (SiC, SiO$_2$, SiN, etc.), in which the NFRHT 
is dominated by surface phonon polaritons (SPhPs) \cite{Mulet2002,Iizuka2015}. There has not been any report
of an extended structure that has heat transfer coefficient exceeding that between two planar polar dielectric
surfaces. In an attempt to tune NFRHT, several calculations of NFRHT between periodic metallic
nanostructures in both 1D \cite{Guerout2012,Dai2015,Dai2016,Messina2016} and 2D \cite{Dai2016b} have 
been reported. These calculations have shown some degree of tunability and a NFRHT enhancement over the 
corresponding material without nanostructuration. However, the reported NFRHT in these structures is still much 
smaller than in the simple case of parallel plates made of polar dielectrics. There have also been
theoretical studies of the NFRHT between photonic crystals and periodic metamaterials made of dielectrics 
\cite{Rodriguez2011,Liu2015,Liu2015b} that show how the radiative properties can be enhanced with respect 
to the bulk counterpart. However, the resulting NFRHTs are again much smaller in comparison with planar
polar dielectrics. 

\begin{figure}[t]
\begin{center} \includegraphics[width=0.9\columnwidth,clip]{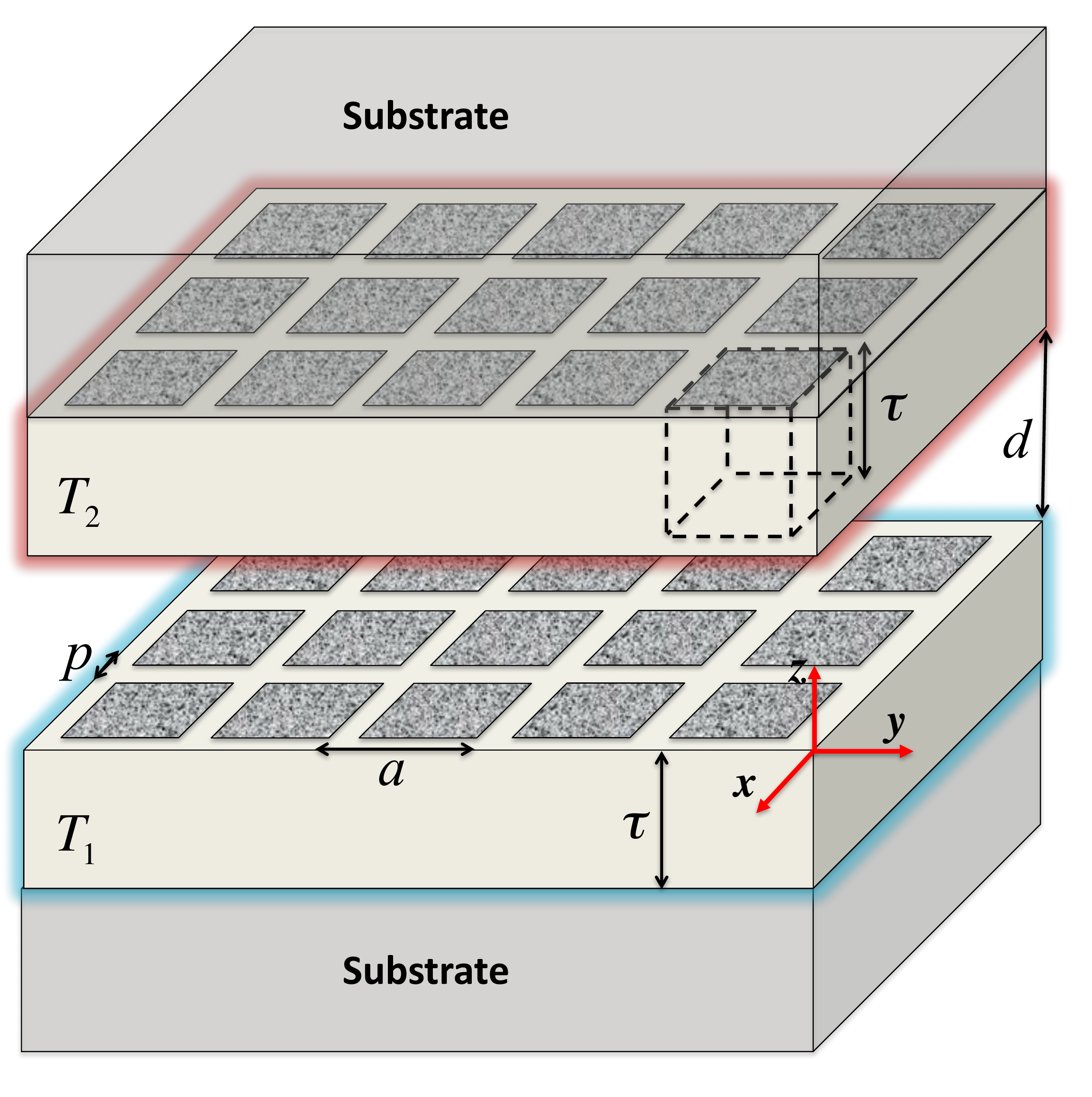} \end{center}
\caption{(Color online) Schematics of two doped-Si metasurfaces made of 2D periodic arrays 
of square holes placed on semi-infinite planar substrates and held at temperatures $T_1$ and $T_2$. 
The key parameters are shown: lattice parameter ($a$), distance between holes ($p$), gap size ($d$), 
and metasurface thickness ($\tau$).}
\label{fig-scheme}
\end{figure}

In this Letter we show that metasurfaces of doped Si (see Fig.~\ref{fig-scheme}) can be used to boost 
NFRHT. Making use of a rigorous coupled wave analysis, we demonstrate that one can design Si 
metasurfaces that not only exhibit a room-temperature NFRHT much larger than that of bulk Si or other 
proposed periodic structures \cite{Liu2015b,Liu2015,Dai2016b}, but they also outperform the best unstructured 
polar dielectric (SiO$_2$). By appropriately choosing the geometrical parameters of the metasurfaces, 
the enhancement over polar dielectrics occurs over a broad range of separations (from 13 nm to 2 $\mu$m). 
The underlying physical mechanisms responsible of this striking behavior are the existence of 
broad-band surface-plasmon polaritons (SPPs) in doped Si, and the ability to tune via nanostructuration 
the dispersion relation of these SPPs that dominate NFRHT in our structure. The predictions of this work 
show the great potential of metasurfaces for the field of NFRHT and can be tested with recent advances 
to measure NFRHT in parallel extended structures \cite{Song2016}.  

The system that we consider is schematically shown in Fig.~\ref{fig-scheme}. It consists of two
identical metasurfaces formed by 2D periodic arrays of square holes drilled in a doped Si layer. 
The metasurfaces are eventually deposited in semi-infinite planar substrates. The geometrical 
parameters of the metasurfaces are the lattice constant $a$, the distance between holes $p$, the 
gap size $d$, and the thickness of the metasurfaces $\tau$, which is equal to the depth of the holes in the
structure. We define the filling  factor of this structure as $f = (a-p)^2/a^2$, which describes the fraction of
vacuum in the structure ($f = 0$ means no holes, while $f=1$ means no Si). We focus on the analysis of the 
heat transfer coefficient, i.e., the linear radiative thermal conductance per unit of area, at 
room temperature (300 K). The dielectric function of doped Si is described within a Drude model 
\cite{Basu2010}: $\epsilon_{\rm Si} (\omega) = \epsilon_{\infty} - \omega^{2}_p/(\omega^2 + i\gamma \omega)$, 
where $\epsilon_{\infty} = 11.7$, $\omega_{p}=711$ meV is the plasma frequency, and $\gamma=61.5$ 
meV is the damping. These values correspond to a doping level of $10^{20}$ cm$^{-3}$. The choice of 
this material and the doping level were motivated by the possibility to sustain SPPs at frequencies 
that can be thermally excited at room temperature. This not possible for very high doping levels, while 
for very low ones the SPPs are not very confined and give a modest contribution to the NFRHT. Our guiding principle 
is the idea that by introducing holes in the Si layers, one can reduce the losses and the effective plasma 
frequency that, in turn, should lead to a redshift of the surface modes. This way, these surface 
modes could be more easily thermally occupied at room temperature leading to an enhancement of the NFRHT. 

To test this idea we have combined the framework of fluctuational electrodynamics (FE) \cite{Rytov1953} 
with a rigorous coupled wave analysis (RCWA) \cite{Caballero2012} to compute NFRHT between periodic 
systems. In the frame of FE, the heat transfer coefficient (HTC) for two arbitrary periodic multilayer 
structures is given by \cite{Bimonte2009}
\begin{equation}
\label{eq-HTC}
h(d,T) = \int^{\infty}_{0} \frac{d \omega}{2\pi} \frac{\partial \Theta(\omega,T)}{\partial T}
\int \frac{d \mathbf k_{\parallel}}{(2\pi)^2} \mathcal{T}(\omega,\mathbf k_{\parallel}) ,
\end{equation}
where $\Theta(\omega,T) = \hbar \omega/ [\exp(\hbar \omega / k_{\rm B}T) -1]$ is the mean energy 
of Planck oscillators at temperature $T$, $\omega$ is the angular frequency, $\mathbf k_{\parallel} 
= (k_x,k_y)$ is the wave vector parallel to the surface planes, and $\mathcal{T}(\omega,\mathbf k_{\parallel})$ 
is the sum over polarizations of the transmission probability of the electromagnetic waves. Note that the 
second integral in Eq.~(\ref{eq-HTC}) is carried out over all possible directions of $\mathbf k_{\parallel}$ 
and it includes the contribution of both propagating waves with $k_{\parallel} < \omega/c$ and evanescent 
waves with $k_{\parallel} > \omega/c$, where $k_{\parallel}$ is the modulus of $\mathbf k_{\parallel}$
and $c$ is the velocity of light in vacuum. Within the RCWA approach, we express the fields in our
periodic system as a sum of plane waves using Bloch theorem. Thus, the transmission function above can be 
obtained by combining scattering matrices of the different interfaces in reciprocal space. In particular, 
if we place the coordinates origin at metasurface 1, the transmission coefficient can be expressed as 
\cite{Bimonte2009}
\begin{equation}
\label{transmission}
\mathcal{T}(\omega,\mathbf k_{\parallel}) = \text{Tr} \left\{ {\hat D}{\hat W_1}{\hat D^\dagger}{\hat W_2} \right\} ,
\end{equation}
where 
\begin{eqnarray}
\hat D & = & ( \hat 1 - \hat S_1 \hat S_2 )^{-1}, \\ 
\hat W_1 & = & \hat \Sigma^{\rm pw}_{-1} - \hat S_1 \hat \Sigma_{-1}^{\rm pw} \hat S_{1}^{\dagger} +
\hat S_{1} \hat \Sigma_{-1}^{\rm ew} - \hat \Sigma_{-1}^{\rm ew} \hat S_{1}^{\dagger} , \\
\hat W_2 & = & \hat \Sigma^{\rm pw}_{+1} - \hat S_{2}^{\dagger} \hat \Sigma_{+1}^{\rm pw} \hat S_{2}+
\hat S_{2}^{\dagger} \hat \Sigma_{+1}^{\rm ew} - \hat \Sigma_{+1}^{\rm ew} \hat S_{2} .
\end{eqnarray}
Here, $\hat S_1 = \hat R_1$ and $\hat S_2 = e^{ik_zd} \hat R_2 e^{ik_zd}$, where $\hat R_1$ and 
$\hat R_2$ are the reflection matrices of the two vacuum-metasurface interfaces and 
$k^2_z = \omega^2/c^2 - k^2_{\parallel}$.
These matrices were computed with the scattering-matrix approach of Ref.~\cite{Caballero2012}.
Moreover, the matrix $\hat \Sigma^{\rm pw(ew)}_{-1(+1)}$ is a projector into the propagating
(evanescent) sector. All these matrices are $2N_g \times 2N_g$ matrices, where $N_g$ is the 
number of reciprocal lattice vectors included in the plane-wave expansions. On the other hand, the 
$\mathbf k$-integral in Eq.~(\ref{eq-HTC}) must be calculated in the interval $(-\pi/a,\pi/a)$ for 
both $k_x$ and $k_y$. A key point in our method is the use of the so-called fast Fourier
factorization when dealing with the Fourier transform of two discontinuous functions in the 
Maxwell equations \cite{Li1997,Caballero2012}. This factorization solves the known convergence 
problems of the RCWA approach.
 
\begin{figure}[t]
\begin{center} \includegraphics[width=0.95\columnwidth,clip]{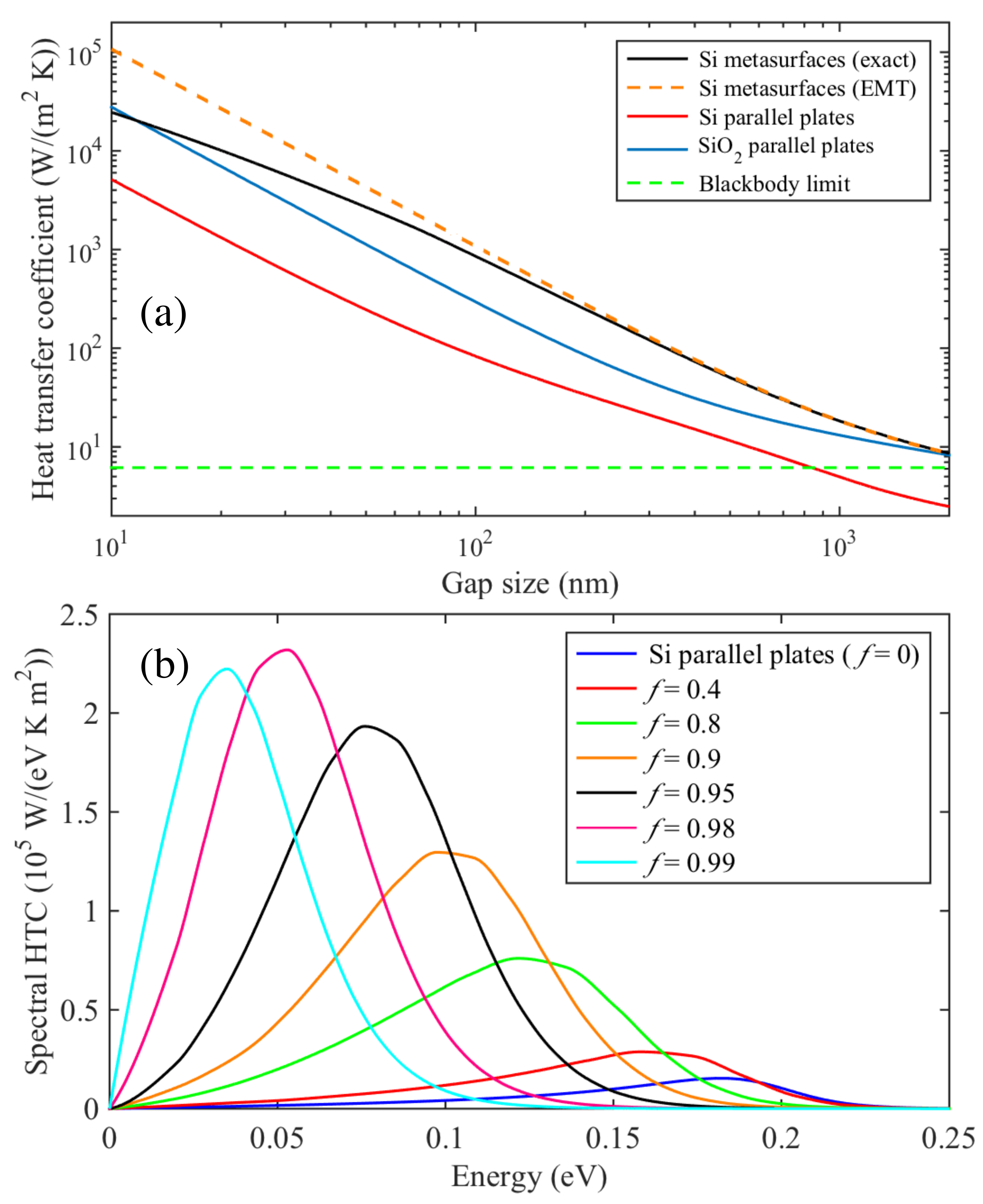} \end{center}
\caption{(Color online) (a) Room-temperature heat transfer coefficient (HTC) as a function of the 
gap size for doped-Si metasurfaces with $a= 50$ nm and $f=0.9$ (black line). For comparison, 
we show the results for the Si metasurfaces computed with effective medium theory (orange dashed
line), SiO$_2$ parallel plates (blue line), and doped-Si parallel plates (red line). The horizontal 
dashed line shows the blackbody limit. (b) Spectral HTC as a function of the photon energy for 
a gap of $d=20$ nm and a lattice constant $a=50$ nm. The different lines correspond to different 
filling factors, as indicated in the legend.}
\label{fig-HTC-spec}
\end{figure}

Let us start the discussion of the results by illustrating the main finding of our work. For
simplicity, we first assume that the Si layer thickness is infinite (no substrate) and below we 
discuss the effect of a finite layer thickness. In Fig.~\ref{fig-HTC-spec}(a) we show the room-temperature 
HTC as a function of the gap size in the near-field regime for two metasurfaces with realistic 
parameters $a=50$ nm and $f=0.9$. This result is compared with the corresponding HTC for two doped-Si 
and two SiO$_2$ parallel plates. As one can see, the NFRHT between the Si metasurfaces is more 
than an order of magnitude larger than the corresponding result for Si plates for a broad range 
of separations, which illustrates the importance of the periodic nanostructuration. More importantly, 
the Si metasurfaces also exhibit a higher HTC than the silica plates in a broad distance range (from 
13 nm to 2 microns), an enhancement that reaches up to a factor 3 for gap sizes of about 100 nm. 
Let us emphasize that our proposed structure exhibits a super-Planckian radiative heat transfer in 
the whole range of gap size as considered in Fig.~\ref{fig-HTC-spec}(a) (see dashed green line).

To get insight into the origin of this NFRHT enhancement due to the nanostructuration, 
we show in Fig.~\ref{fig-HTC-spec}(b) the spectral HTC of the metasurfaces for a gap
$d=20$ nm, a lattice constant $a=50$ nm, and for different filling factors. As one can observe,
the maximum of the spectral HTC is shifted towards lower frequencies upon increasing the size of
the holes from 0.2 eV for $f=0$ up to around 0.05 eV for $f=0.98$. Notice also that the HTC
(the integral of these spectral functions) also increases drastically with the filling factor reaching
a maximum at $f \approx 0.98$. These results illustrate the high tunabilibity of NFRHT in 
metasurfaces.  

\begin{figure}[t]
\begin{center} \includegraphics[width=1.0\columnwidth,clip]{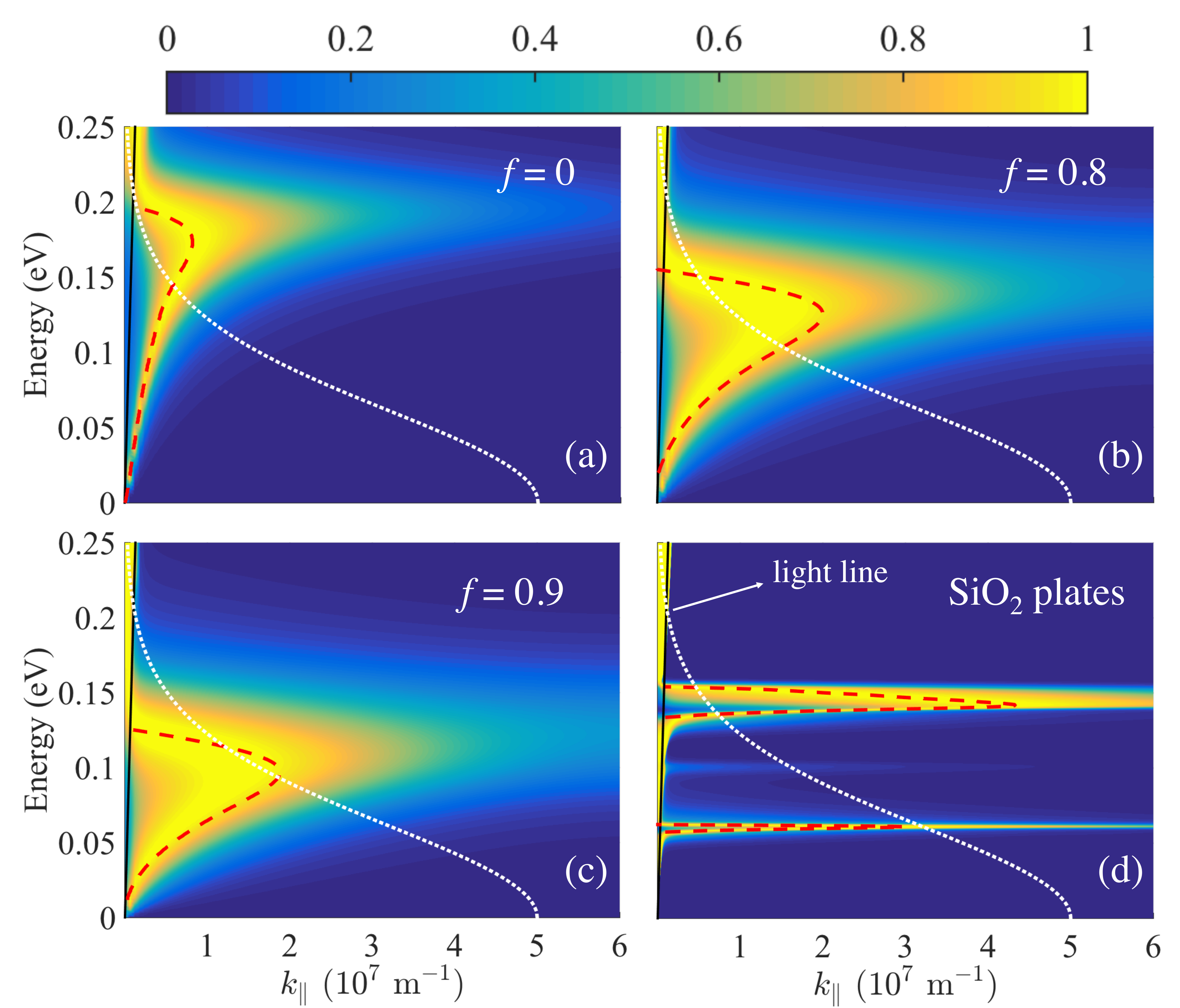} \end{center}
\caption{(Color online) Transmission function of p-polarized waves along the $x$-direction for 
a gap $d=20$ nm. (a-c) Si metasurfaces with lattice parameter $a=50$ nm. The different panels 
correspond to various filling factors. The red dashed lines correspond to the dispersion relation 
of the SPPs as obtained from Eq.~(\ref{eq-disprel}) and the black lines are light lines in vacuum. 
(d) The same as in the other panels but for two SiO$_2$ parallel plates. In this case, the red
dashed line corresponds to the dispersion relation of the SPhPs in this structure. The white 
dotted lines in all panels correspond the thermal occupation function $\partial \Theta(\omega,T)/ 
\partial T$, see Eq.~(\ref{eq-HTC}), in arbitrary units.}
\label{fig-trans}
\end{figure}

The origin of the redshift in the spectral HTC can be understood with an analysis of the 
frequency and parallel wave vector dependence of the transmission $\mathcal{T}(\omega,\mathbf k_{\parallel})$. 
Such a dependence is displayed in Fig.~\ref{fig-trans} for p-polarized waves, which dominate the NFRHT.
In particular, we show the transmission along the $x$-direction [$\mathbf k_{\parallel} = (k_x,0)$],
see Fig.~\ref{fig-scheme}, for $d=20$ nm, $a = 50$ nm, and different filling factors. As one 
can see in Fig.~\ref{fig-trans}(a) for the case of two Si parallel plates ($f=0$), the transmission
maxima resemble the dispersion relation of a surface electromagnetic mode. As we shall show below, it corresponds 
to a cavity SPP mode that emerges from the hybridization of the SPP modes of the two vacuum-Si interfaces.
Notice in particular that the transmission maxima lie to the right of the light line, which indicates
that these modes correspond to evanescent waves (both in vacuum and inside Si). As the filling factor 
increases, we find that the transmission maxima shift towards lower energies, see Fig.~\ref{fig-trans}(b-c), 
which is consistent with our observation above about the spectral HTC. Indeed, the frequency at which
the maxima of the spectral HTC occur, see Fig.~\ref{fig-HTC-spec}(b), corresponds exactly to the position
at which the transmission maxima fold back towards the light line. The reason is that in that frequency 
region the transmission is not only maximum, but also $k_{\parallel}$ takes the largest value, maximizing 
thus the density of photonic modes. Again, NFRHT is dominated by cavity SPPs which arise from the hybridization
of the SPP modes at the extended surface of both metasurfaces.
 
To confirm that cavity SPPs are indeed responsible for the NFRHT in our structure, we have analyzed
their dispersion relation. For this purpose, we have made used of an effective medium theory (EMT) \cite{Liu2013b}.
Within this theory our periodic metasurfaces can be modeled as uniaxial materials with a diagonal 
permittivity tensor: $\hat {\epsilon} = \mbox{diag}(\epsilon_{\rm o}, \epsilon_{\rm o}, \epsilon_{\rm e})$, 
where the subindex o and e denote the ordinary and extraordinary optical axis, respectively. The components 
of the dielectric tensor are given by \cite{Liu2013b}
\begin{equation}
\label{eq-eo-ee}
\epsilon_{\rm o} = \epsilon_{\rm Si} \frac{\epsilon_{\rm Si} (1-f)+1+f }{ \epsilon_{\rm Si}(1+f)+1-f},
\;\; \epsilon_{\rm e} =  f+(1-f)\epsilon_{\rm Si}.
\end{equation}
In such an anisotropic system, light propagates along the ordinary and extraordinary axes with 
perpendicular components of the wave vector given by $k^2_{z,{\rm o}} = \epsilon_{\rm o} \omega^2/c^2 - 
k^2_{\parallel}$ and $k^2_{z,{\rm e}} = \epsilon_{\rm o} \omega^2/c^2 - k^2_{\parallel} \epsilon_{\rm o}/
\epsilon_{\rm e}$, respectively. Within this uniaxial approximation, the SPP dispersion relation is given 
by the solution of the following equation \cite{Moncada2015}
\begin{equation}
\label{eq-disprel}
e^{ik_zd} = \pm \left( \frac{k_{z,{\rm e}} - \epsilon_{\rm o} k_z}
{k_{z,{\rm e}} + \epsilon_{\rm o} k_z} \right),
\end{equation}
where let us recall that $k^2_z = \omega^2/c^2 - k^2_{\parallel}$. In the electrostatic 
limit ($k_{\parallel} \gg \omega/c$), this equation leads to the following dispersion relation 
for the cavity SPPs
\begin{equation}
\label{eq-k-SPP}
k_{\rm SPP}(\omega) = \frac{1}{d} \ln \left( \pm \frac{\epsilon_{\rm o}(\omega) - 
\sqrt{\epsilon_{\rm o}(\omega)/\epsilon_{\rm e}(\omega)}}{\epsilon_{\rm o}(\omega) + 
\sqrt{\epsilon_{\rm o}(\omega)/ \epsilon_{\rm e}(\omega)}} \right) .
\end{equation}
We have solved Eq.~(\ref{eq-disprel}) and plotted the corresponding dispersion relations in 
Fig.~\ref{fig-trans} (red dashed lines). As it can be seen, these dispersion relations nicely 
coincide with the transmission maxima for the whole range of filling factors. This unambiguously 
demonstrates that cavity SPPs dominate the NFRHT in our system. Moreover, this shows that NFRHT 
is drastically enhanced upon increasing the filling factor because the surfaces modes shift to 
lower frequencies, which increases their thermal occupation at room temperature. This is illustrated
in Fig.~\ref{fig-trans} where we show as white dotted lines the frequency dependence of the  
thermal factor $\partial \Theta(\omega,T)/ \partial T$ that determines the occupation of these
surface modes.

The enhancement of NFRHT in our metasurfaces over polar dielectrics like SiO$_2$ can also be
understood with an analysis of the transmission. As we show in Fig.~\ref{fig-trans}(d), the 
transmission between two silica parallel plates is dominated by SPhPs \cite{Mulet2002}, whose
dispersion relation is given by Eq.~(\ref{eq-k-SPP}) with $\epsilon_{\rm o} = \epsilon_{\rm e}
= \epsilon_{\rm SiO_2}$ \cite{Song2015a}. Although theses surface modes exhibit larger 
$k_{\parallel}$-values than the SPPs in our Si metasurfaces and therefore larger photonic
density of states, they are restricted to rather narrow frequency regions corresponding
to the two Reststrahlen bands in this material. Thus, the larger extension in frequency of 
the SPPs in the Si structures is one of the key factors that leads to the higher NFRHT in these 
metasurfaces.

The fact that EMT nicely describes the position of the maxima of the spectral NFRHT raises 
the question of whether this theory can accurately describe all the results in this type of 
metasurfaces. This is actually not the case because EMT assumes that the geometrical features 
are much smaller than some of the relevant physical length scales of the problem. In our case, where 
there is a considerable damping, the natural lateral scale is set by the propagation length of the
cavity SPP wavelengths, $1/(2\mbox{Im}\{k_{\rm SPP}\})$. For frequencies close to the folding 
back of the dispersion relation, which are the ones that dominate the spectral HTC, this propagation 
length can obtained from Eq.~(\ref{eq-k-SPP}). Thus for instance, for the structure analyzed in 
Fig.~\ref{fig-HTC-spec}(a), Eq.~(\ref{eq-k-SPP}) predicts that the SPP propagation length for the 
frequency of the spectral HTC maximum becomes of the order of the lattice parameter for $d \approx 100$ 
nm. Thus, the EMT is expected to fail below this gap size. To confirm this idea, we have computed 
the HTC in this structure within the EMT using the formalism for anisotropic planar systems of 
Ref.~[\onlinecite{Moncada2015}] and the result is shown in Fig.~\ref{fig-HTC-spec}(a) as a
dashed orange line. As one can see, the EMT indeed fails for gaps below the SPP propagation length. 
This analysis illustrates the need of an exact approach, like our RCWA, to accurately predict the 
NFRHT in these metasurfaces.

\begin{figure}[!t]
\begin{center} \includegraphics[width=0.95\columnwidth,clip]{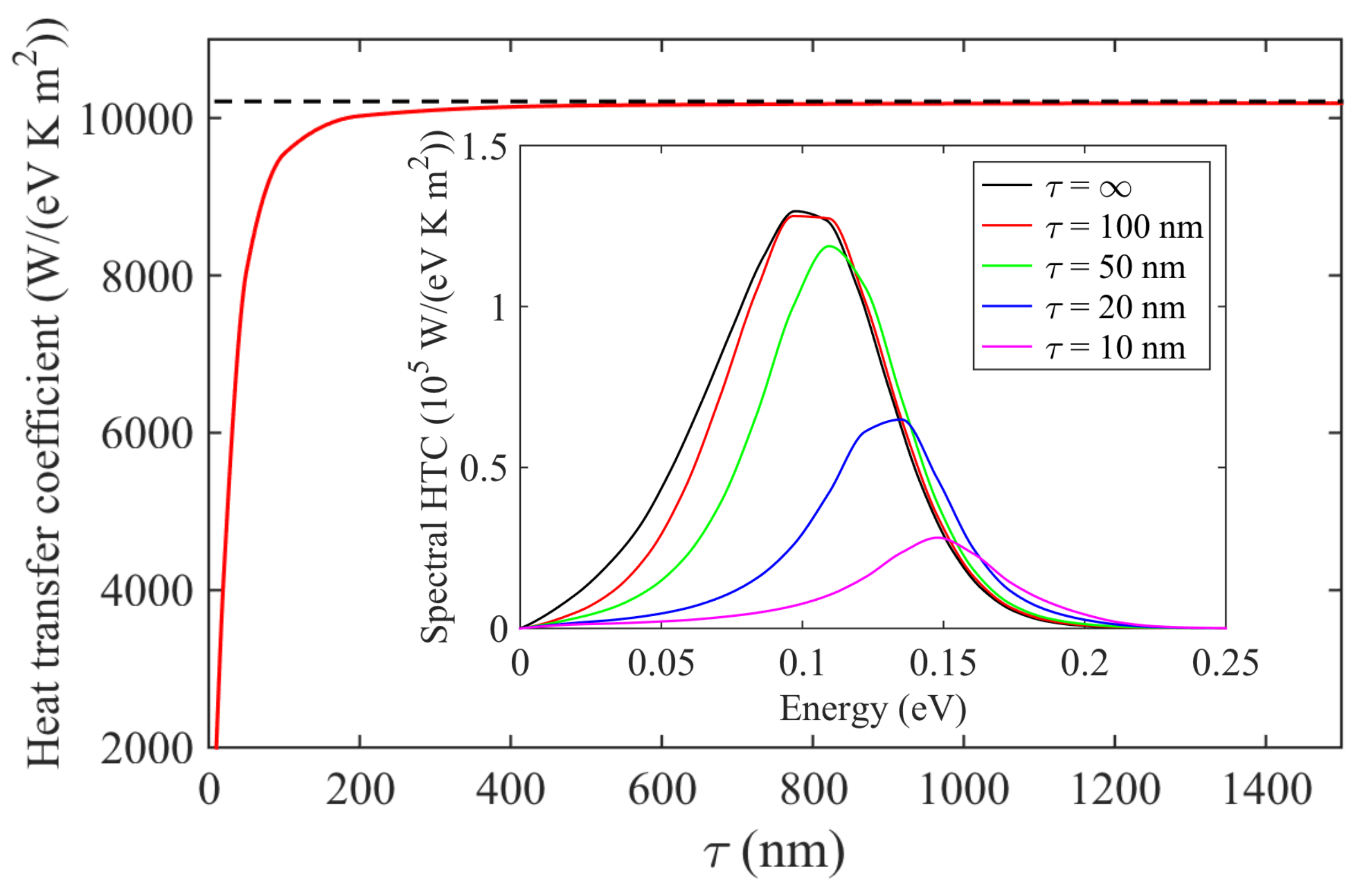} \end{center}
\caption{(Color online) Heat transfer coefficient as a function of the thickness of the
periodically patterned Si layers for $d=20$ nm, $a=50$ nm, and $f=0.9$. The dashed line 
corresponds to $\tau = \infty$. The inset shows the spectral HTC as a function of energy 
for several metasurface thicknesses.}
\label{fig-tau}
\end{figure}

It is worth stressing that the periodic structures discussed above truly behave as metasurfaces.
This can be understood as follows. From Eq.~(\ref{eq-k-SPP}) we can estimate the penetration
depth of the cavity SPPs, which in the electrostatic limit is given by $1/(2\mbox{Re}\{k_{\rm SPP}\})$.
Thus, we see that this penetration depth diminishes as the gap size is reduced, which 
implies that the NFRHT is dominated by the surface of the periodic structures. Thus for instance,
for $d=20$ nm, $a=50$ nm, and $f=0.9$ the penetration depth estimated from Eq.~(\ref{eq-k-SPP}) is 
about 27 nm for the frequency of the spectral HTC maximum. Thus, one expects that periodic 
structures with Si layers thicker than this penetration depth behave exactly in the same way.  
To test this idea we have computed the HTC as a function of the thickness of the periodic Si layers, 
$\tau$, assuming that the substrate underneath is also made of (unstructured) doped Si. In 
Fig.~\ref{fig-tau} we show the results for this thickness dependence for the case mentioned above. 
Notice that when the layer thickness becomes larger than the gap, which is comparable to the SPP
penetration depth, the HTC quickly tends to the result for a semi-infinite structure. This behavior
is illustrated in the inset of Fig.~\ref{fig-tau} with the corresponding spectral HTCs. Thus, we
can conclude that our periodic structures effectively behave as true metasurfaces as long as the 
thickness of the periodically patterned part is larger than the gap size. 

In summary, we have proposed a novel mechanism to further enhance NFRHT with the use of Si metasurfaces, 
which is based on the broad spectral bandwidth and the high tunability of the SPPs that dominate NFRHT 
in these structures. We have shown that by an appropriate choice of the geometrical parameters,
these metamaterials can exhibit room-temperature near-field radiative heat conductances higher 
than any existent or proposed structure. The fabrication of these metasurfaces is feasible with 
the state-of-the-art nanolithography \cite{Manfrinato2013} and our predictions could be tested with 
the recent developments in the measurement of NFRHT in parallel extended structures \cite{Song2016}.

This work has been funded by ``La Caixa" Foundation, the Spanish MINECO under Contracts No.\ 
FIS2014-53488-P and MAT2014-53432-C5-5-R, the Comunidad de Madrid (Contract No.\ S2013/MIT-2740),
and the European Research Council (ERC-2011-AdG Proposal No.\ 290981). J.C.C.\ and V.F.-H. thank 
the DFG and SFB 767 for sponsoring their stay at the University of Konstanz (J.C.C.\ as Mercator Fellow). 
S.F.\ acknowledges the support of the Global Climate and Energy Project (GCEP) at Stanford University, and 
the U.S.\ Department of Energy ``Light-Material Interactions in Energy Conversion" Energy Frontier Research
Center under Grant No.\ DE-SC0001293.


\end{document}